\documentclass[aps,12pt]{revtex4}

\usepackage{graphicx}
\usepackage{graphics}
\usepackage{color}
\usepackage{amsfonts}

\def\beq{\begin{equation}}
\def\eeq{\end{equation}}
\def\bea{\begin{eqnarray}}
\def\eea{\end{eqnarray}}

\begin{document}

\title{Algebraic approach to directed stochastic avalanches}
\author{B. L. Aneva~$^{\dag^1}$ and J. G. Brankov~$^{\dag^2\dag^3}$ \email{brankov@theor.jinr.ru}}

\address{$^{\dag^1}$~Institute for Nuclear Research and Nuclear Energy, Bulgarian
Academy of Sciences, 1784 Sofia, Bulgaria}

\address{$^{\dag^2}$~Bogoliubov Laboratory of Theoretical Physics, Joint
Institute for Nuclear Research, 141980 Dubna, Russia}

\address{$^{\dag^3}$~Institute of Mechanics, Bulgarian
Academy of Sciences, 1113 Sofia, Bulgaria}

\begin{abstract}
A two-dimensional directed stochastic sandpile model
is studied analytically with the use of directed Abelian
algebras recently introduced by Alcaraz and V. Rittenberg
[Phys. Rev. E {\bf 78}, 041126 (2008)].
Exact expressions for the probabilities
of all possible toppling events which follow the
transfer of arbitrary number of particles  to a site in the
stationary configuration are derived. A description of the virtual-time
evolution of directed avalanches on two dimensional lattices is suggested.
Due to intractability of the general problem, the algebraic approach
is applied only to the solution of the special cases of directed deterministic
avalanches and trivial stochastic avalanches describing simple random walks
of two particles. The study of
these cases has clarified the role of each particular kind of
toppling in the process of avalanche growth.
In the general case of the quadratic directed algebra  we have
determined exactly the maximum possible values of: (1) the current
of particles at any given
moment of virtual time and (2) the occupation number (`height') of each site at
any moment of time.



\end{abstract}

\maketitle

\section{Introduction}

Sandpile models, introduced in 1987 by Bak, Tang and Wiesenfeld
(BTW), have drawn a lot of attention as the simplest systems which
describe Self-Organized Criticality with intrinsic avalanche-like
dynamics resembling the one observed in nature \cite{BTW}. Despite
their simplicity and the great efforts invested in their solution,
a rigorous derivation of the critical exponents describing the
stationary state of the isotropic BTW models is still lacking.
However, the establishment of the Abelian property of the particle
topplings in the critical height models enhanced their analytical
tractability \cite{D90}. A number of important characteristics of
the stationary state have been rigorously derived \cite{PKI,IP,D}.
Next, the deterministic directed sandpiles (DDS) were introduced
and analytically solved \cite{DR}. It became evident that they
belong to a special universality class with exactly known critical
exponents.

Isotropic, as well as directed, sandpile models with stochastic dynamics were
introduced too \cite{M,SV,V}. The numerical evaluation of the critical
exponents of the stochastic directed sandpiles (SDS) has shown that they
belong to a still different universality class \cite{V,SV}.

\begin{figure}
\center
\includegraphics[width=100mm]{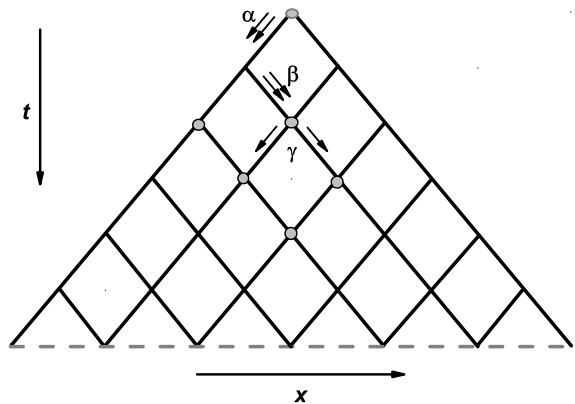}
\caption{Schematic representation of the rotated by $\pi/4$ square
lattice and the directed toppling rules. The bottom boundary of
the lattice is open.}\label{F1}
\end{figure}

Recent extensive Monte Carlo simulations, performed by Alcaraz and
Rittenberg \cite{AR} on the rotated by $\pi /4$ square lattice,
see Fig.~\ref{F1}, have indicated that the two dimensional
directed stochastic sandpiles belong to a universality class with
$\sigma_{\tau} = 1.780 \pm 0.005$. If the estimated error bars are
correct, then this result is in contradiction with the analytical
prediction $\sigma_{\tau} = 1.75$ \cite{PB,KMT}, as well as with
the previous numerical estimates \cite{SV}. Therefore,
reexamination of the critical exponents of directed stochastic
avalanches becomes important. An attempt in that direction was
undertaken in \cite{Bu}.

A new approach to the analytical treatment of directed avalanches
has been suggested by Alcaraz and Rittenberg \cite{AR}. It is
based on the study of directed Abelian algebras (DAAs) on
two-dimensional acyclic lattices. The quadratic algebra suggested
for the lattice shown in Fig.~\ref{F1} acts in the bulk as
\begin{equation}
a^2_{i,j} = \alpha [\mu a^2_{i+1,j} +(1- \mu)a_{i,j}a_{i+1,j}]+
(1-\alpha)[\mu a^2_{i,j+1} +(1- \mu)a_{i,j}a_{i,j+1}],
\label{alAR}
\end{equation}
where $a_{i,j}$ is the generator attached to each site $(i,j)$ of
the lattice. Here the labeling of the lattice sites is such that
$i-1$ is the distance (in lattice spacings) from the right
boundary, and $j-1$ is the distance from the left boundary. Thus,
the nearest neighbors of site $(i,j)$ in the direction of
propagation (downwards) are $(i+1,j)$ (the left neighbor) and
$(i,j+1)$ (the right neighbor).

Equation (\ref{alAR}) describes toppling of particles from an
unstable site $(i,j)$ which involves the following stochastic
events. (1) Two particles topple with probability $\mu$: they both
go to the left (right) nearest neighbor in the direction of
propagation with probability $\alpha \mu$ (resp., $(1- \alpha)
\mu$). (2) One particle topples and the other remains at the same
site with probability $1- \mu$: the toppled particle goes to the
left (right) nearest neighbor with probability $\alpha (1- \mu)$
(resp., $(1 - \alpha)(1- \mu)$).

\section{The stochastic directed quadratic algebra}

Here we shall study a SDS on the rotated square lattice, see Fig. \ref{F1},
with more simple and convenient for theoretical investigation
toppling rules, a particular case of which was considered in \cite{PB} and \cite{KMT}.
According to these rules, any unstable site relaxes to a stable configuration
(with at most one particle) through a succession of two-particle topplings:
the two particles are
transferred to the left (right) nearest neighbor in front
with probability $\alpha$ (resp., $\beta$), or one goes left and the other
goes right with probability $\gamma = 1 -\alpha - \beta$. Each lattice
site can emit only an even number of particles but can receive any
number of them. Therefore, one can readily classify the sites with
respect to their effect on the flux of particles \cite{PB,KMT}. A site
that receives an even number of particles emits the same number of
them, hence, it does not change the flux and is called {\it passive}.
A site that receives an odd number $2n+1$ of particles is {\it active}, since
if empty it retains one particle and emits the remaining even part $2n$ of them
({\it negatively active}), while if occupied it emits the total number of $2n+2$
particles, i.e. increases the flux by one unit ({\it positively active}).
It was shown that the critical state of the above SDS is a product measure
with average particle density $\rho = 1/2$.

Next, we find it
convenient to label the sites so that the first coordinate $i$ is the
integer time step $\tau$, and the second coordinate $j$ numbers the sites which
can be visited by the avalanche at time $\tau =i$ in the horizontal (spatial)
direction. Thus, the lattice
sites form the triangular array $\mathcal{L} =\{(i,j): j=1,2,\dots, i,\; i=1,2,\dots, T\}$,
where $T$ is the size of the lattice in the temporal direction.
In the above notation, the quadratic algebra we study reads
\begin{equation}
a^2_{i,j} = \alpha a^2_{i+1,j} + \beta a^2_{i+1,j+1}+
\gamma a_{i+1,j}a_{i+1,j+1}.
\label{oural}
\end{equation}
Here it is assumed that sites $(i,j)$ do not belong to the open boundary
of the lattice $\partial \mathcal{L} =\{i=T,\; j=1,2, \dots, T\}$.
There are as many algebraic relations (\ref{oural})
as sites in the lattice. With the sites lying on the open boundary
one associates generators satisfying the following $T$ relations
(see Eq. (71) in \cite{AR})
\begin{equation}
a^2_{T,j} = 1,  \quad j=1,2,\dots , T.
\label{boundary}
\end{equation}

In the case under study the critical stationary state of the system is
(see Eq. (73) in \cite{AR} at $\mu =1$)
\begin{equation}
\Phi_{1,T} = \prod_{i=1}^T \prod_{j=1}^i \frac{1+ a_{i,j}}{2}.
\label{ststate}
\end{equation}
This corresponds to a product measure with equal probability of
having a site vacant or occupied by just one particle. One can
show by finite induction that
\begin{equation}
a_{i,j}\Phi_{1,T} = \Phi_{1,T}, \quad (i,j)\in \mathcal{L}.
\label{eigenstate}
\end{equation}

Avalanches will always be started by dropping particles on site $(1,1)$
until it becomes unstable. If we introduce a restriction of the stationary state
to the time interval from $i=\tau$ to $i=T$,
\begin{equation}
\Phi_{\tau,T} = \prod_{i=\tau}^T \prod_{j=1}^i \frac{1+ a_{i,j}}{2},
\label{reststate}
\end{equation}
the beginning of an avalanche will be described as
\begin{equation}
a^2_{1,1}\Phi_{2,T} = (\alpha a^2_{2,1} + \beta a^2_{2,2}+
\gamma a_{2,1}a_{2,2})\frac{1+ a_{2,1}}{2}\frac{1+ a_{2,2}}{2} \Phi_{3,T}.
\label{beginava}
\end{equation}
One can read from here the obvious fact that the avalanche may stop at the
second time-step $\tau =2$ with probability $\gamma/4$: if the two particles on
the initial site go to different neighbors (with probability $\gamma$) and
both of these neighbors are empty (with probability 1/4).

In order to compute the probabilities of larger avalanches, one has to determine
the effect of an arbitrary number of particles piled up on a
given site at time $\tau$ on the restriction of the stationary state to the
interval $[\tau, T]$. In particular, one has to compute for any integer $n$
the product $a^n_{i,j}(1+ a_{i,j})/2$. Since the result depends on the parity
of $n$, we consider separately $n=2p$ even,
\begin{equation}
a^{2p}_{i,j}\frac{1+ a_{i,j}}{2}= \frac{1}{2}\sum_{k=0}^{2p}C_k^{(2p)}(1+ a_{i,j})
a_{i+1,j}^{2p-k}a_{i+1,j+1}^k,
\label{neven}
\end{equation}
and $n=2p+1$ odd,
\begin{equation}
a^{2p+1}_{i,j}\frac{1+ a_{i,j}}{2}= \frac{1}{2}\sum_{k=0}^{2p}C_k^{(2p)}
a_{i,j}a_{i+1,j}^{2p-k}a_{i+1,j+1}^k + \frac{1}{2}\sum_{k=0}^{2p+2}C_k^{(2p+2)}
a_{i+1,j}^{2p+2-k}a_{i+1,j+1}^k.
\label{nodd}
\end{equation}

In the former case, when an even number $2p$ of particles comes to a
stationary site $(i,j)$, all the $2p$ particles topple: $2p-k$ to the left
and $k$ to the right with probability $C_k^{(2p)}$, $k=0,1,\dots, 2p$. At
that the state of the site $(i,j)$ remains unchanged.

In the latter case, when an odd number $2p+1$ of particles comes to a
stationary site $(i,j)$, the result depends on the occupation number of
that site. If the site is empty, only $2p$ particles topple: $2p-k$ to the left
and $k$ to the right with probability $C_k^{(2p)}$, $k=0,1,\dots, 2p$, and
one particle remains on that site. If the site is occupied, all the
$2p+2$ particles will topple: $2p+2-k$ to the left
and $k$ to the right with probability $C_k^{(2p+2)}$, $k=0,1,\dots, 2p+2$, and
the site remains empty. However, the net state of the site $(i,j)$
will not change, because the stationary probabilities of being empty or
occupied by one particle are equal.

By deriving recurrent relations for the coefficients $C_k^{(2p)}$,
\begin{equation}
C_k^{(2p+2)} = \alpha C_k^{(2p)} + \beta C_{k-2}^{(2p)}+ \gamma C_{k-1}^{(2p)},
\quad k=2,\dots, p,
\label{recurr}
\end{equation}
and solving
them under the initial conditions: $C_0^{(0)}=1$, $C_0^{(2)}=\alpha$, $C_1^{(2)}=\gamma$,
and $C_2^{(2)}=\beta$, we obtain
\begin{equation}
C_k^{(2p)}(\alpha, \beta,\gamma) = \sum_{m=0}^{[k/2]}\frac{p!}{(k-2m)!m!(p+m-k)!}
\alpha^{p+m-k} \beta^m \gamma^{k-2m},
\quad k=0,1,\dots, p,
\label{Ck2p}
\end{equation}
where $[m]$ denotes the entire part of the real number $m$. Together with the
left-right symmetry property
$$C_k^{2p}(\alpha, \beta,\gamma)=C_{2p-k}^{2p}(\beta,\alpha,\gamma),$$
Eq. (\ref{Ck2p}) completely defines the coefficients $C_k^{2p}(\alpha, \beta,\gamma)$,
$k=0,1,\dots, 2p$. Here are their explicit expressions for $p=2,3$:
\begin{eqnarray}
C_0^{(4)} &=& \alpha^2 ,\quad C_1^{(4)} = 2\alpha \gamma, \quad C_2^{(4)} = 2\alpha \beta +
\gamma^2, \quad C_3^{(4)} = 2\beta \gamma, \quad C_4^{(4)} = \beta^2 \nonumber \\
C_0^{(6)} &=& \alpha^3 ,\quad C_1^{(6)} = 3\alpha^2 \gamma, \quad C_2^{(6)} = 3\alpha^2 \beta +
3 \alpha \gamma^2, \quad C_3^{(6)} = 6\alpha \beta \gamma +\gamma^3, \nonumber \\
&&C_4^{(6)} = 3\alpha \beta^2 + 3 \beta \gamma^2, \quad C_5^{(6)} = 3\beta^2 \gamma,
\quad C_6^{(6)} = \beta^3.
\label{p2p3}
\end{eqnarray}
Note that $C_k^{(2p+1)} = 0$, $k=0,1, \dots, 2p+1$.

In order to describe the size distribution of the avalanche, we
need to know only the number of particles transferred from
time-step $\tau$ to time-step $\tau +1$. The probability with
which that number vanishes for the first time is the
probability of having avalanches with time duration $\tau$.
Another distribution we are interested in is the probability
distribution of having avalanches with a given total number of
toppled particles, which measures the ``size'' of an avalanche. In
all these cases we are not interested in the configuration changed
by the avalanche. Following Alcaraz and Rittenberg \cite{AR}, we
use the symbol $\hat{=}$ to denote expressions in which all the
generators of the algebra left behind the front of the avalanche
are replaced by unity. For example,
\begin{equation}
a^{2p}_{i,j}\frac{1+ a_{i,j}}{2}\; \hat{=} \sum_{k=0}^{2p}C_k^{(2p)}
a_{i+1,j}^{2p-k}a_{i+1,j+1}^k, \label{eventop}
\end{equation}
\begin{equation}
a^{2p+1}_{i,j}\frac{1+ a_{i,j}}{2}\; \hat{=}\; \frac{1}{2}\sum_{k=0}^{2p}C_k^{(2p)}
a_{i+1,j}^{2p-k}a_{i+1,j+1}^k + \frac{1}{2}\sum_{k=0}^{2p+2}C_k^{(2p+2)}
a_{i+1,j}^{2p+2-k}a_{i+1,j+1}^k.
\label{oddtop}
\end{equation}

Let $P(n_1,\dots, n_\tau |\tau)$ denote the probability that at time $i=\tau$
the sites $\{(\tau,1),(\tau,2), \dots, (\tau,\tau)\}$
have occupation numbers $\{n_1, n_2, \dots, n_\tau \}$, respectively. The total number
of particles at that moment may range from $0$ to
some finite $n_\mathrm{max}(\tau)$. Now, the virtual-time evolution of avalanches
on the lattice $\mathcal L$ is described by
\begin{equation}
a^2_{1,1}\Phi_{2,T}\; \hat{=} \sum_{n=0}^{n_\mathrm{max}(\tau)}
\left[\sum_{n_1 +\cdots + n_\tau = n}
P(n_1,\dots, n_\tau |\tau)\prod_{k=1}^\tau a^{n_k}_{\tau,k}\right]\Phi_{\tau+1,T} \;,\quad
\tau = 2,\dots, T-1.
\label{evolava}
\end{equation}
Of course, the avalanche continues from moment $\tau$ to moment $\tau +1$ only if there is at
least one $n_k \geq 2$. In this case
the monomial $\prod_{k=1}^\tau a^{n_k}_{\tau,k}$ represents a possible distribution
of the particles in the row $\tau$, and the flux of particles that hits the next row
$\tau +1$ is obtained by applying formulas (\ref{eventop}) or (\ref{oddtop}) to each
$a^{n_k}_{\tau,k}$ with $n_k \geq 2$. The
span of the avalanche front at the moment $\tau$ is from $j_\mathrm{min}(\tau)$ to
$j_\mathrm{max}(\tau)$, where $j_\mathrm{min}(\tau)$ ($j_\mathrm{max}(\tau)$) is the
leftmost (rightmost) unstable site.

The analysis of the avalanche evolution from Eq. (\ref{evolava})
for large times $\tau$ seems untractable problem. In order to get
some insight about the role of the different toppling processes,
we  pass to the consideration of two extreme cases of the algebra
(\ref{oural}): $\alpha = \beta = 0, \gamma =1$ and $\alpha = \beta
= 1/2, \gamma =0$.

\subsection{Directed deterministic avalanches}

\begin{figure}[t]
\center
\includegraphics[width=100mm]{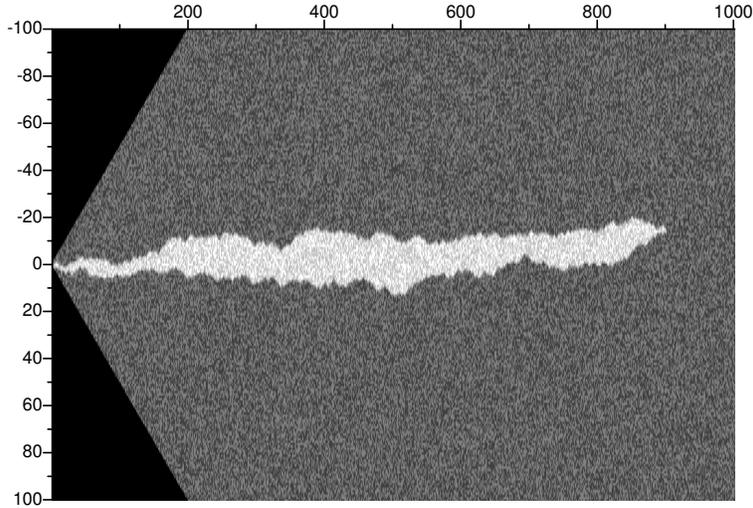}
\caption{Trace of a directed deterministic avalanche of duration
901 time steps.}\label{F2}
\end{figure}

When $\alpha = \beta = 0$, hence, $\gamma =1$, the
evolution of the avalanche becomes deterministic, because the two
toppling particles always go to different nearest neighbors in
front. Thus, the left and right boundaries of the unstable
avalanche cluster perform simple random walks, see Fig.~\ref{F2}.
In this case the coefficients (\ref{Ck2p}) reduce to
\begin{equation}
C_k^{(2p)}(0, 0,1) = \delta_{k,p},
\label{g}
\end{equation}
and relations (\ref{eventop}), (\ref{oddtop}) simplify to
\begin{eqnarray}
&& a^{2p}_{i,j}\frac{1+ a_{i,j}}{2}\; \hat{=}\; a_{i+1,j}^{p}a_{i+1,j+1}^p, \nonumber \\
&& a^{2p+1}_{i,j}\frac{1+ a_{i,j}}{2}\; \hat{=}\; \frac{1}{2}a_{i+1,j}^{p}a_{i+1,j+1}^p +
\frac{1}{2}a_{i+1,j}^{p+1}a_{i+1,j+1}^{p+1}.
\label{equivg}
\end{eqnarray}

Now we shall prove that the virtual-time evolution of
the deterministic avalanche, described by (\ref{evolava}) with the
toppling rules (\ref{equivg}), simplifies drastically. The avalanche
front becomes ``compact'' (without gaps of stable sites) and almost ``flat'':
the unstable sites may have only two or three particles. In the
first step of the evolution this is trivially true, since
\begin{equation}
a_{1,1}^2\frac{1+a_{2,1}}{2}\frac{1+a_{2,2}}{2} =
\frac{1}{4}(a_{2,1}a_{2,2}+a_{2,1}^2 a_{2,2}+a_{2,1} a_{2,2}^2 +a_{2,1}^2 a_{2,2}^2)
\hat{=} \frac{1}{4}(1+ a_{2,1}^2 + a_{2,2}^2+ a_{2,1}^2 a_{2,2}^2).
\label{first}
\end{equation}
It is seen that with probability 1/4 the avalanche stops and, whenever
it continues, the unstable sites at $\tau =2$ are occupied by exactly two
particles. In the next step we obtain
\begin{eqnarray}
&& a_{1,1}^2\prod_{i=2}^3 \prod_{j=1}^i \frac{1+a_{i,j}}{2}\;
\hat{=} \frac{1}{4}(1+ a_{3,1} a_{3,2} + a_{3,2}a_{3,3}+ a_{3,1}a_{3,2}^2a_{3,3})
\prod_{j=1}^3 \frac{1+a_{3,j}}{2}\nonumber \\
&&\hat{=} \frac{1}{32}(12+ 2a_{3,1}^2+ 5a_{3,2}^2 + 2a_{3,3}^2+ a_{3,2}^3+ 3a_{3,1}^2a_{3,2}^2
+ 3a_{3,2}^2 a_{3,3}^2+a_{3,1}^2a_{3,2}^3\nonumber\\&& + a_{3,2}^3a_{3,2}^2+
a_{3,1}^2a_{3,2}^2a_{3,3}^2+ a_{3,1}^2a_{3,2}^3a_{3,3}^2).
\label{second}
\end{eqnarray}
Obviously, the set of unstable sites remains compact and the occupation numbers
of the unstable sites equal only 2 or 3. Having established these properties
for $\tau =3$, we prove now that they persist for $\tau +1$.

According to our assumption, at some $\tau$ the general term of the
operator products in the right-hand side of (\ref{evolava}) has the form
\beq
a_{\tau,1}^{n_1}\cdots a_{\tau,p}^{n_{p}}a_{\tau,p+1}^{n_{p+1}}
\cdots a_{\tau,p+q}^{n_{p+q}}
a_{\tau,p+q+1}^{n_{p+q+1}}\cdots a_{\tau,\tau}^{n_{\tau}}, \label{3}
\eeq
where $n_1,\dots ,n_p$ and $n_{p+q+1},\dots ,n_{\tau}$ take values $0,1$
and $n_{p+1},\dots ,n_{p+q}$ equal 2 or 3. This term describes an avalanche
with front spanning $q$ adjacent unstable sites. Since each unstable site
emits only two particles, after the substitution
\beq
a_{\tau, i}\hat{=}1, \quad a_{\tau, i}^2=a_{\tau +1,i}a_{\tau +1, i+1}, \quad
a_{\tau, i}^3\hat{=}a_{\tau +1,i}a_{\tau +1, i+1}, \qquad i\in \{1,2,\dots, \tau\},
\label{2} \eeq
the terms that are relevant for generation of the avalanche evolution at time step
$\tau+1$ take the form
\begin{eqnarray}
&&\prod_{k=1}^\tau a_{\tau,k}^{n_k} \Phi_{\tau+1,T} \hat{=}
\prod_{k=p+1}^{p+q}a_{\tau +1,k}a_{\tau +1,k+1}\left[\prod_{i=1}^{\tau +1}
\frac{1+ a_{\tau +1,i}}{2}\right] \Phi_{\tau+2,T} \nonumber \\
&& = \left[\prod_{i=1}^{p}\frac{1+ a_{\tau +1,i}}{2}\right] \frac{a_{\tau +1,p+1}+ a_{\tau +1,p+1}^2}{2}
\left[\prod_{k=p+2}^{p+q}\frac{a_{\tau +1,k}^2+ a_{\tau +1,k}^3}{2}\right]\nonumber \\
&& \times \frac{a_{\tau +1,p+q+1}+ a_{\tau +1,p+q+1}^2}{2}\Phi_{\tau+2,T}\nonumber \\
&&\hat{=}\left(\frac{1}{2}+ \frac{1}{2}a_{\tau +1,p+1}^2\right)
\left[\prod_{k=p+2}^{p+q}a_{\tau +1,k}^2\right]\left(\frac{1}{2}+ \frac{1}{2}a_{\tau +1,p+q+1}^2\right)
\Phi_{\tau+2,T}.
\label{3}
\end{eqnarray}

From the above expressions one can tell that: (1) the avalanche front remains compact;
(2) the unstable sites can have height 2 or 3 only; (3) if the hitting avalanche front is of
length $q$ at time $\tau$, in the next moment of time $\tau+1$ it  can:
(a) shrink to $q-1$ with probability 1/4; (b) remain of the same length $q$
with probability 1/2; (c) expand to length $q+1$ with probability 1/4.  The latter property
reflects the fact that the left and right boundary of the compact
avalanche front undergo independently simple random walks one step to the
left and one step to the right. At that, a step to the left neighbor in front does not
change the distance from the left boundary, while a step to the right increases it by
one unit. Denoting by $P(L, \tau)$ the probability of an avalanche to have front of length
at time $\tau$, one obtains the recurrence relation
\beq  P(L, \tau+1)= \frac{1}{4}P(L-1, \tau)+ \frac {1}{2}P(L, \tau) +
\frac{1}{4}P(L+1,\tau). \label {6}
\eeq

\subsection{Simple random walks}

\begin{figure}[h]
\center
\includegraphics[width=100mm]{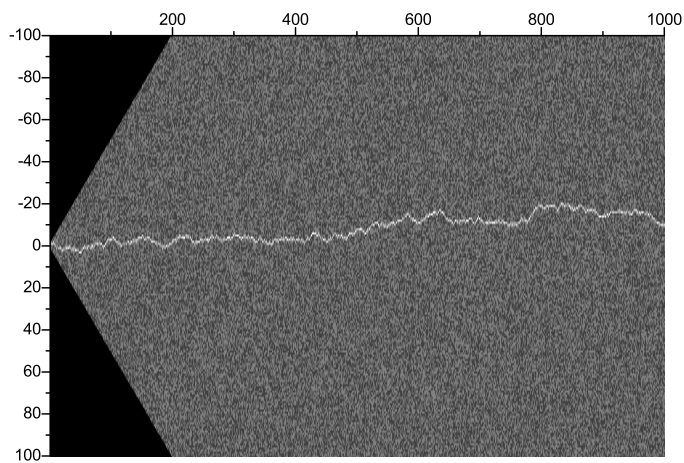}
\caption{Trace of a directed stochastic avalanche of the simple
random walk type.}\label{F3}
\end{figure}

In the case $\alpha = \beta = 1/2, \gamma =0$ we obtain avalanches
carrying exactly two particles throughout the whole lattice, the
trace of which represents an unbiased simple random walk, see
Fig.~\ref{F3}. The coefficients (\ref{Ck2p}) become
\begin{equation}
C_{2m}^{(2p)}(1/2,1/2,0) = 2^{-p}{p \choose m},
\quad C_{2m+1}^{(2p)}(1/2,1/2,0)=0, \quad m=0,1,\dots, p-1,
\label{ab}
\end{equation}
and relations (\ref{eventop}), (\ref{oddtop}) take the form
\begin{eqnarray}
&& a^{2p}_{i,j}\frac{1+ a_{i,j}}{2}\; \hat{=}\; 2^{-p}\sum_{m=0}^p {p \choose m}
a_{i+1,j}^{2(p-m)}a_{i+1,j+1}^{2m}, \nonumber \\
&& a^{2p+1}_{i,j}\frac{1+ a_{i,j}}{2}\; \hat{=}\; 2^{-p-1}\sum_{m=0}^p {p \choose m}
a_{i+1,j}^{2(p-m)}a_{i+1,j+1}^{2m} +
2^{-p-2}\sum_{m=0}^{p+1} {p+1 \choose m}a_{i+1,j}^{2(p+1-m)}a_{i+1,j+1}^{2m}.\qquad
\label{equivab}
\end{eqnarray}

Now the quadratic algebra (\ref{oural}) has two terms \beq
a_{i,j}^2=\frac {1}{2}a_{i+1,j}^2+\frac{1}{2}a_{i+1,j+1}^2
\label{7} \eeq which act on the stationary state independently,
without changing it. We shall prove that the front of each
avalanche now has just one unstable site occupied by 2 or 3
particles. The avalanches perform a simple random walk and end up
at the open boundary of the system at $\tau =T$.

In the initial step of the avalanche evolution, one has
\begin{eqnarray}
&&a_{1,1}^2\frac{1+a_{2,1}}{2}\frac{1+a_{2,2}}{2} =
\frac{1}{8}(a_{2,1}^2 + a_{2,1}^3+a_{2,1}^2 a_{2,2}+a_{2,1}^3
a_{2,2} +a_{2,2}^2 +a_{2,1}a_{2,2}^2+a_{2,2}^3
+a_{2,1}a_{2,2}^3)\nonumber \\&& \hat{=}\;\frac{1}{2}a_{2,1}^2 +
\frac{1}{2}a_{2,2}^2 = \frac{1}{4}(a_{3,1}^2 + 2a_{3,2}^2+
a_{3,3}^2). \label{first2}
\end{eqnarray}
It is seen that with equal probability 1/2 the two particles of
the avalanche go to the left or to the right nearest-neighbor
ahead. In the next step, the row $\tau =2$ emits again exactly 2
particles, which are distributed according to the unbiased simple
random walk probabilities. After hitting the stationary
distribution of the row $\tau =3$, the avalanche continues in the
same way:
\begin{eqnarray}
&& a_{1,1}^2\prod_{i=2}^3 \prod_{j=1}^i \frac{1+a_{i,j}}{2}\;
\hat{=} \frac{1}{4}(a_{3,1}^2 + 2a_{3,2}^2+ a_{3,3}^2)
\prod_{j=1}^3 \frac{1+a_{3,j}}{2}\nonumber \\
&&\hat{=} \frac{1}{4}(a_{3,1}^2 + 2a_{3,2}^2+ a_{3,3}^2)=
\frac{1}{8}(a_{4,1}^2 + 3a_{4,2}^2+ 3a_{4,3}^2 +a_{4,4}^2).
\label{second2}
\end{eqnarray}
Thus, up to $\tau =3$, the front of each avalanche contains
exactly one unstable site with occupation number equal to 2 or 3
only. The emitted particles from layer to layer are exactly 2.
Having established these properties for $\tau =3$, we prove now
that they persist for $\tau +1$.

According to our assumption, at some $\tau$ the general term of
the operator products in the right-hand side of (\ref{evolava})
has the form
\begin{eqnarray}
&&a_{\tau,j}^2\Phi_{\tau+1,T}=\frac{1}{2}(a_{\tau+1,j}^2+a_{\tau+1,j+1}^2)
\left[\prod_{i=1}^{\tau+1}\frac{1+ a_{\tau +1,i}}{2}\right]
\Phi_{\tau+2,T}\nonumber \\
&& = \left[\prod_{i=1}^{j-1}\frac{1+ a_{\tau +1,i}}{2}\right]
\frac{a_{\tau +1,j}^2+ a_{\tau +1,j}^3}{4}
\left[\prod_{i=j+1}^{\tau+1}\frac{1+ a_{\tau +1,i}}{2}\right]\Phi_{\tau+2,T}
\nonumber \\
&&+\left[\prod_{i=1}^{j}\frac{1+ a_{\tau +1,i}}{2}\right]
\frac{a_{\tau +1,j+1}^2+ a_{\tau +1,j+1}^3}{4}
\left[\prod_{i=j+2}^{\tau+1}\frac{1+ a_{\tau +1,i}}{2}\right]
\Phi_{\tau+2,T}\nonumber \\
&&\hat{=}\frac{1}{4}(a_{\tau +2,j}^2+2a_{\tau+2,j+1}^2
+a_{\tau+2,j+2}^2) \Phi_{\tau+2,T}. \label{3rw}
\end{eqnarray}

For general values of $\alpha$ and $\beta = 1-\alpha$, the
algebraic form of the evolution at any virtual time corresponds to
processes where the two particles from site $(1,1)$ pass through
the whole lattice along trajectories with probability distribution
corresponding to the biased simple random walk.

\subsection{Exact results for the stochastic avalanche}

In this section we present exact results concerning some important
extremal characteristics of the directed stochastic avalanches
which obey the general algebra (\ref{oural}). We derive the
maximum value $I_{\rm max}(\tau)$ of the current of particles and
the maximum height $h_{\rm max}(\tau,j)$ at any site $(\tau,j)$,
$j=1,\dots, \tau$, for any given moment of virtual time $\tau$.

\subsubsection{The maximum current at virtual time $\tau$}

\begin{figure}[h]
\center
\includegraphics[width=100mm]{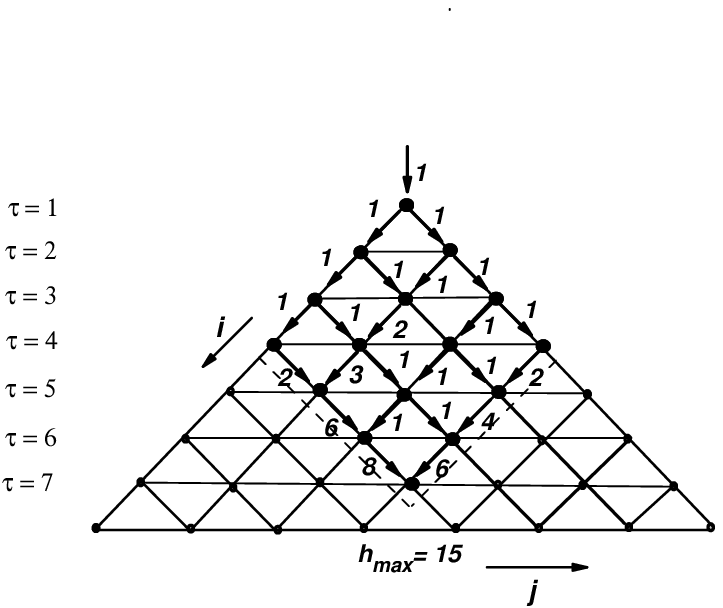}
\caption{Schematic illustration of an avalanche leading to a
maximum unstable height at the central site of an odd-$\tau$ row.
The integers besides the arrows indicate the number of particles
transferred in the corresponding direction}\label{F4}
\end{figure}

At $\tau =1$ the value of $I_{\rm max}(1)=2$ is the initial
condition for an avalanche to start. At $\tau =2$ the value of
$I_{\rm max}(2)=4$ is attained when both sites $(2,1)$ and $(2,2)$
are positively active, i.e., they are occupied and receive one
particle each in the toppling $a_{1,1}^2 \rightarrow \gamma
a_{2,1}a_{2,2}$. In general, the maximum possible current of
particles $I_{\rm max}(\tau)$ at virtual time $\tau$ is attained
whenever the row $\tau -1$ has emitted the maximum possible
current $I_{\rm max}(\tau -1)$, and the maximum possible number of
sites in the row $\tau$ are positively active, i.e., occupied and
visited by an odd number of particles. Since, due to the algebra
(\ref{oural}), the current is always even, the latter quantity
obviously depends on the parity of $\tau$. Examples of such
topplings between virtual times $\tau$ = 1,2,3, and 4 are shown in
Fig. \ref{F4}.

First we prove the following proposition.

{\it Proposition I.} The maximum current of particles that leave
the row $\tau$ is given by \bea I_{max}(\tau)&=&\frac
{\tau^2+1}{2}+1,  \qquad \mathrm{for} \quad \tau \; \mathrm{odd};
 \label{Imaxodd}\\
I_{max}(\tau)&=&\frac{\tau^2}{2}+2, \qquad \qquad \mathrm{for}
\quad \tau \; \mathrm{even}. \label{Imaxeven} \eea

{\it Proof.} The proof contains two important ingredients. The
first one is to find avalanches which  transfer the maximum
possible number of particles from row to row. To ensure most
favorable conditions, we consider stable configurations in which
all the sites in the rows $1\leq t \leq \tau$ are occupied. The
second one is the derivation of recurrence relation between
$I_{\rm max}(\tau)$ and $I_{\rm max}(\tau-2)$. The solutions of
this recurrence, separately for odd and even $\tau$, yield the
desired proof.

We begin with proving that any even-$\tau$ row can have all of its
sites positively active, therefore it can emit $I_{\rm
max}(\tau-1)+ \tau$ particles. Such an event occurs as a result of
the following sequence of topplings. First, each of the $\tau/2$
odd-numbered sites in the row $\tau-1$, that is $\{(\tau-1,2m-1),
m=1,\dots ,\tau/2\}$, emits two particles in a process \beq
a_{\tau -1,2m-1}^2 \rightarrow \gamma a_{\tau,2m-1}a_{\tau,2m},
\quad m=1,\dots ,\tau/2, \label{allg} \eeq so that each site of
the row $\tau$ receives exactly one particle. This is possible
because, for $\tau \geq 3$, the unstable configuration of the
layer $\tau-1$ contains maximum $I_{\rm max}(\tau -2) +\tau -1$
particles: $I_{\rm max}(\tau -2)$  received from the preceding
layer $\tau -2$, and $\tau -1$ particles at the sites in the
stationary configuration of the layer $\tau -1$. After the
completion of the topplings (\ref{allg}), in the row $\tau-1$
there will remain $I_{\rm max}(\tau -2) -1$ particles. The even
part of that number, namely $I_{\rm max}(\tau -2) -2$ particles,
can be transferred to row $\tau$ in arbitrary portions of even
numbers and one particle will remain in the row $\tau -1$. Thus,
the total number of particles accumulated in the unstable
configuration of row $\tau$ is $I_{\rm max}(\tau -2) +2\tau -2$
and all of them are partitioned in even portions on it sites.
Therefore,
\begin{equation}
I_{\rm max}(\tau)= I_{\rm max}(\tau-1)+ \tau =I_{\rm max}(\tau -2) +2\tau -2,
\quad \tau \geq 4 \; \mathrm{even}.
\label{Irecureven}
\end{equation}

Next we prove that an odd-$\tau$ row can have at most $\tau-1$
positively active sites. Obviously, not all $\tau$ sites can be
active, because the received number of particles $I_{\rm
max}(\tau-1)$ is always even and cannot be distributed in odd
portions among the odd number $\tau$ of sites. To prove that
exactly one site can remain passive, we consider the processes
(\ref{allg}) for $m=1,\dots ,(\tau -1)/2$ which transfers exactly
one particle to each of the first $\tau-1$ sites in the row
$\tau$. The last unstable site $(\tau -1, \tau-1)$ in the previous
row can emit either an even number of particles to the site
$(\tau,\tau)$, thus leaving it passive, or it can send odd
portions of particles to each of the sites $(\tau,\tau-1)$ and
$(\tau,\tau)$. In the latter case the result will be
$(\tau,\tau-1)$ passive and $(\tau,\tau)$ positively active. The
existence of just one passive site in the row $\tau$ will not
change when the remaining unstable sites in the row $\tau -1$
topple even portions of particles to sites in $\tau$. Hence, an
odd row $\tau$ can emit maximum $I_{\rm max}(\tau) =I_{\rm
max}(\tau-1) +\tau -1$ particles. Since $\tau-1$ is even, by the
previous argument we have $I_{\rm max}(\tau-1)= I_{\rm
max}(\tau-2) +\tau -1$. Therefore,
\begin{equation}
I_{\rm max}(\tau)= I_{\rm max}(\tau-1)+ \tau -1 =I_{\rm max}(\tau -2) +2\tau -2,
\quad \tau \geq 3\; \mathrm{odd}.
\label{Irecurodd}
\end{equation}

Summarizing, independently of the parity of $\tau$, we obtain the
recurrence relation \beq I_{\rm max}(\tau)-I_{\rm
max}(\tau-2)=2\tau -2, \qquad \tau >2. \label{Irecgen} \eeq

To solve the above recurrence, we note that for odd moments of
virtual time, $\tau = 2n-1$, $n=1,2,\dots$, it yields
\begin{equation}
\sum_{k=2}^n [I_{\rm max}(2k-1)- I_{\rm max}(2k -3)] \equiv I_{\rm
max}(2n-1)- I_{\rm max}(1) = \sum_{k=2}^n(4k -4) = 2n^2 -2n.
\label{sumeven}
\end{equation}
By taking into account the initial condition $I_{\rm max}(1)=2$
and substituting $n=(\tau +1)/2$, we prove the first part of
Proposition I.

For an even $\tau = 2n$, $n=1,2,\dots$, we obtain from the recurrence
(\ref{Irecgen})
\begin{equation}
\sum_{k=2}^n [I_{\rm max}(2k)- I_{\rm max}(2k -2)] \equiv I_{\rm
max}(2n)- I_{\rm max}(2) = \sum_{k=2}^n(4k -2) = (2n)^2 -2.
\label{sumeven}
\end{equation}
Then, by taking into account the initial condition $I_{\rm
max}(2)=4$ and substituting $n=\tau/2$, we complete the proof of
Proposition I.

\subsubsection{The maximum height at a site at time $\tau$}

The proof of the results presented here uses the following definition.

{\it Definition.} For each site $(\tau,j)$, $2\leq \tau \leq T$,
$1\leq j \leq \tau$, we define a {\it basin of attraction} of particles
as the set of all the preceding sites, which can send
particles to that site by means of an avalanche obeying the algebra
(\ref{oural}).

The principle of establishing the maximum  possible occupation number
of a site is to consider avalanches which realize the maximum possible
current from layer to layer within the basin of attraction of the given
site. Since, by intuition (at least, when $\alpha =\beta$), the largest
occupation numbers for
algebras (\ref{oural}) are reached at the middle of the row,
we consider first the central sites $(\tau,(\tau+1)/2)$ for $\tau$ odd,
and $(\tau,\tau/2 -1)$, $(\tau,\tau/2 +1)$ for $\tau$ even.

{\it Proposition II.} The maximum height attained at the central site of
a row $\tau$ is:

(1) At odd moments of virtual time $\tau =2n-1$, the maximum
height at sites $(2n-1,n\pm p)$, $p=0,1,2,\dots ,n-1$, is
\begin{equation}
h_{\rm max}(2n-1,n\pm p)= \left\{
\begin{array}{lll}
n(n-1)-p(p-1)+3, &\mathrm{for}\quad n-p  & \mathrm{even} \\
n(n-1)-p(p+1)+3, &\mathrm{for}\quad n-p  & \mathrm{odd}.
\end{array}\right.
\label{lochodd}
\end{equation}

The global maximum of the height is reached at the central site
$(2n-1,n)$ and equals
\begin{equation}
h_{\rm max}(\tau,(\tau+1)/2)= (\tau^2-1)/4 +3.
\label{hodd}
\end{equation}

(2) At even moments of virtual time $\tau =2n$, the maximum height
at each of the sites $(2n,n-p)$, $(2n,n+p+1)$, $p=0,1,2,\dots
,n-1$, is
\begin{equation}
h_{\rm max}(2n,n-p)=h_{\rm max}(2n,n+p+1)= \left\{
\begin{array}{lll}
n^2-p^2+3, &\mathrm{for}\quad n-p  & \mathrm{even} \\
n^2-p^2-2(p-1), &\mathrm{for}\quad n-p  & \mathrm{odd}.
\end{array}\right.
\label{locheven}
\end{equation}

The global maximum of the height is reached at each of the central
site $(2n,n)$, $(2n,n+1)$ and equals
\begin{equation} h_{\rm max} (2n,n)=h_{\rm max}(2n,n+1)= \left\{
\begin{array}{lll}
\tau^2/4 +3, &\mathrm{for}\quad \tau/2  & \mathrm{even} \\
\tau^2/4 +2, &\mathrm{for}\quad \tau/2  & \mathrm{odd}.
\end{array}\right.
\label{heven}
\end{equation}

{\it Proof.} We consider again the most favorable stable
configuration, which is the fully occupied basin of attraction of
the given site. In the proof of Proposition I, when counting the
number of particles carried
downstream by an avalanche, we established the following results:

(i) An avalanche passing from even row to the next odd row can
transfer ahead at most {\it all but one} of the particles in the
fully occupied stable configuration of the odd row.

(ii) There exists an avalanche which on passing from odd row to
the next even row can transfer ahead {\it all} of the particles in
the fully occupied stable configuration of the even row.

The above statements were proved in the case when all the sites of a
pair of subsequent rows can participate
in the avalanche. In the present consideration we need analogous
results for avalanches spreading only in the basin of attraction
of a given site. Thus we encounter the problem of transfer of
particles between segments of subsequent rows and we prove first
the following stronger results.

{\it Lemma 1.} Consider segments of two consecutive rows which
obey the condition that each site in one of the segments has at
least one nearest neighbor in the other one. Then:

(a) If the number of sites in the second segment is odd,
avalanches can transfer downstream at most {\it all but one} of
the particles in the fully occupied stable configuration of that
segment.

(b) If the number of sites in the second segment is even, there
exists an avalanche which can transfer downstream {\it all} the
particles in the fully occupied stable configuration of that
segment.

{\it Proof.} There are three possible types of configurations of
the two segments which satisfy the conditions of the lemma. With
respect to the change in number of sites on passing downstream
from segment to segment, these can be classified as follows:

\noindent (1) With increasing number of sites,
\begin{eqnarray}
&& (i,j+1), (i,j+2), \dots, (i,j+m-1) \nonumber \\
&& (i+1,j+1), (i+1,j+2), \dots, (i+1,j+m-1), (i+1,j+m);
\label{incr} \end{eqnarray}
(2) With equal number of sites,
\begin{eqnarray}
&& (i,j+1), (i,j+2), \dots, (i,j+m-1), (i,j+m) \nonumber \\
&& (i+1,j+1), (i+1,j+2), \dots, (i,j+m-1), (i+1,j+m); \label{eql}
\end{eqnarray}
\noindent or
\begin{eqnarray}
&& (i,j), (i,j+1), \dots, (i,j+m-1) \nonumber \\
&& (i+1,j+1), (i+1,j+2), \dots, (i,j+m-1), (i+1,j+m). \label{eqr}
\end{eqnarray}
(3) With decreasing number of sites,
\begin{eqnarray}
&& (i,j),(i,j+1), \dots, (i,j+m-1), (i,j+m) \nonumber \\
&& (i+1,j+1), (i+1,j+2), \dots, (i,j+m-1), (i+1,j+m). \label{decr}
\end{eqnarray}
Due to the left-right symmetry of the lattice, when both $\alpha
>0$ and $\beta >0$, it suffices to consider only one of the realizations
of case (2), say, the configuration (\ref{eqr}).

Note that in our notation the number of sites in the second
segment is always $m$. With regard to the parity of $m$, each of
the above cases splits into two subcases: (a) $m$ odd, and (b) $m$
even.

(a) Let $m$ be odd. No avalanche can turn all the sites of the
second segment $i+1$ into positively active ones, because every
avalanche transfers downstream an even number of particles which
cannot be distributed in odd portions among an odd number of
sites. However, there exist an avalanche which can make all but
one of the occupied sites in the segment $i+1$ positively active.
An example of such an avalanche includes the topplings
\begin{equation}
a^2_{i,k} \rightarrow \gamma a_{i+1,k}a_{i+1,k+1}, \label{top1}
\end{equation}
with $k= j+1, j+3,\dots, j+m-2$. As a result, in each of the
configurations (1)-(3), all the sites of the second segment,
except the last one $(i+1,j+m)$, receive exactly one particle. All
the remaining unstable sites in the uper segment $i$ can emit even
portions of particles to their neighbors in segment $i+1$. Thus
all $m-1$ sites $(i+1,k)$, $k= j+1, j+2,\dots, j+m-1$, become
positively active, only the last site $(i+1,j+m)$ remains passive
or, possibly, stable in the cases (\ref{incr}) and (\ref{eqr}).

(b) Let $m$ be even. An avalanche, which turns all the sites of
the second segment into positively active ones, is constructed as
follows. After the topplings (\ref{top1}) with $k= j+1, j+3,\dots,
j+m-1,$ all the sites in the lower segment $i+1$ receive exactly
one particle. Then, the remaining unstable sites in the upper
segment $i$ transfer even portions of particles to their nearest
neighbors in the downstream segment $i+1$. The existence of at
least one such neighbor is ensured by the conditions of the lemma.

Now we turn back to the proof of Proposition II. Since the details
of the analysis depend on the parity of $\tau$, the different
cases are considered separately.

(1) Consider first the case of odd $\tau = 2n-1$, $n=2,3,\dots$.
Then the basin of attraction of the central site $(2n-1,n)$ is the
square with vertices at sites $(1,1)$, $(n,1)$, $(n,n)$ and
$(2n-1,n)$, see Fig. \ref{F4}. Provided an avalanche has started,
the total number of particles in the basin of attraction is
$n^2+1$. However, not all of these particles can be delivered to
the target site $(2n-1,n)$, because, before reaching that site,
the avalanche has to pass through $N_{\rm odd}(2n-1)=n-2$
intermediate rows with odd number of sites, leaving a particle at
each of them. Therefore, \beq h_{\rm max} (2n-1,n) = n^2+1
-(n-2)=n^2-n+3 = (\tau^2-1)/4 +3. \label{odd} \eeq

Consider next a shift by $p=1,2,\dots, n-1$ sites to the left or
to the right of the central site $(2n-1,n)$. Such a shift changes
the basin of attraction from the square $n\times n$ to a rectangle
$(n+p)\times (n-p)$. Thus, the number of particles in the fully
occupied stable configuration of the basin of attraction of the
sites $(2n-1,n\pm p)$ decreases to $n^2 -p^2$. Remarkably, the new
rectangular basin of attraction contains $2p+1$ segments (from row
$n-p$ to row $n+p$) which have equal number $n-p$ of sites. Thus,
the maximum possible number of particles transferred by an
avalanche to the sites $(2n-1,n\pm p)$ depends on the parity of
$n-p$, as well as on the parity of $n$.

(1a) When $n-p$ is even, all the particles on the $2p+1$ central
even-length segments can be transferred downstream by an avalanche
(see Lemma 1). Excluding the initial and final sites, there remain
$2(n-p-2)$ unequal-length segments, half of which contain odd
number of sites. Hence, the maximum number of particles that
can occupy sites $(2n-1,n\pm p)$ is
\begin{equation}
h_{\rm max}(2n-1,n\pm p)= n(n-1) -p(p-1)+3 \qquad \mathrm{for}
\quad n-p \quad \mathrm{even}. \label{oddeven}
\end{equation}

Hence, in the case of maximal shift $p=n-2$ ($p$ must be odd) one
obtains
$$h_{\rm max}(2n-1,2)= h_{\rm max}(2n-1,2n-2)= 4n-3=2\tau -1.$$

\begin{figure}[t]
\center
\includegraphics[width=100mm]{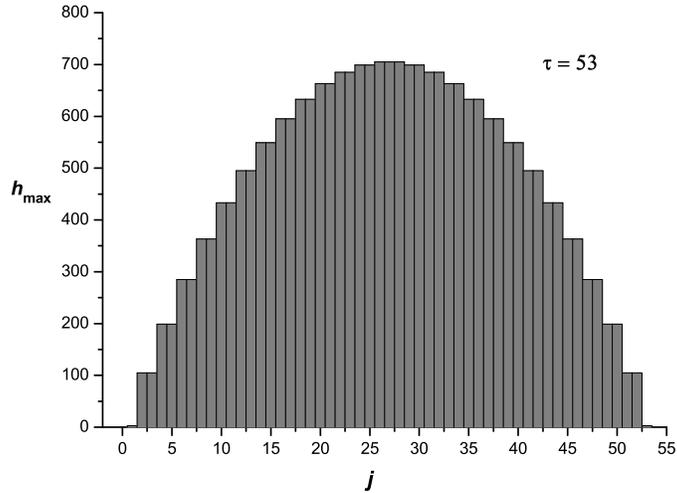}
\caption{Distribution of the maximum possible unstable height at a
site at virtual-time moment $\tau =53$. Since $(\tau +1)/2$ is
odd, the maximum is attained at the three central sites, although
in different avalanches, see Discussion.} \label{F53}
\end{figure}

(1b) When $n-p$ is odd, all the avalanches will leave one particle
on the $2p+1$ central odd-length segments (see Lemma 1). Excluding
the initial and final sites, there again remain $2(n-p-2)$
unequal-length segments. However, now between the initial (final)
site and the central $2p+1$ odd-length segments there are two
series of segments each of which begins and ends up with an
even-length segment (the first one from 2 to $n-p-1$ and the
second one from $n+p+1$ to 2). Hence, the remaining odd-length
segments are $n-p-3$. Therefore, the maximum number of particles
that can occupy sites $(2n-1,n\pm p)$ is
\begin{equation}
h_{\rm max}(2n-1,n\pm p)= n(n-1) -p(p+1)+3 \qquad \mathrm{for}
\quad n-p \quad \mathrm{odd}. \label{oddodd}
\end{equation}
Note that at $p=0$ expressions (\ref{oddeven}) and (\ref{oddodd})
yield the same result (\ref{odd}) for the global maximum at time
$\tau =2n-1$, irrespectively of the parity of $n$. Remarkably, for
$n$ odd the substitution $p=1$ in (\ref{oddeven}) shows that all
three central sites have the same height $h_{\rm max} (2n-1,n\pm
1)=h_{\rm max}(2n-1,n)$. This case is illustrated in
Fig.~\ref{F53} for $n=27$.

\begin{figure}[h]
\center
\includegraphics[width=100mm]{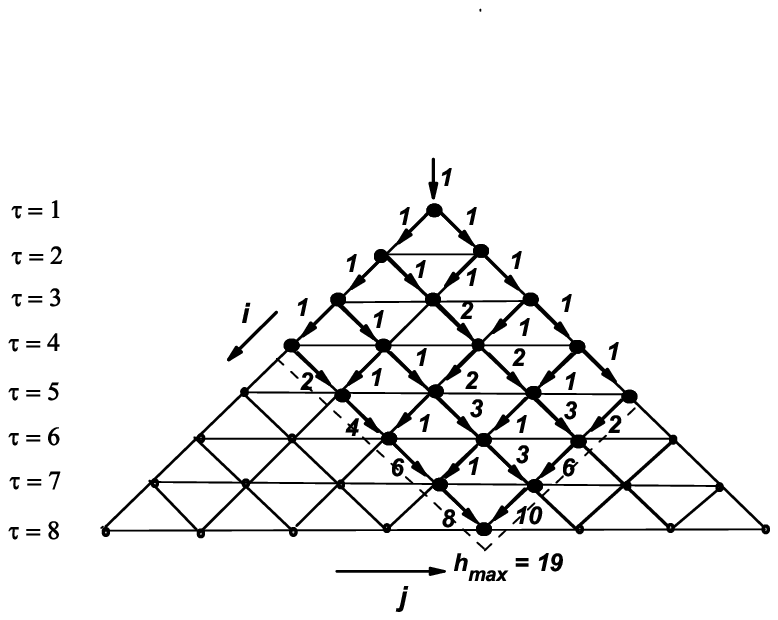}
\caption{Schematic illustration of an avalanche leading to a
maximum unstable height at a central site of an even-$\tau$ row
when $\tau/2$ is even. The integers besides the arrows indicate
the number of particles transferred in the corresponding
direction}\label{F5}
\end{figure}

(2) Consider now the case of even $\tau = 2n$, $n=2,3,\dots$. Due
to the symmetry of the lattice, each of the two central sites
$(2n,n)$ and $(2n,n+1)$ has the same maximum occupation number.
For definiteness, consider the basin of attraction of the site
$(2n,n+1)$. It represents a rectangle $n \times (n+1)$ with
vertices at sites $(1,1)$, $(n,1)$, $(n+1,n+1)$ and $(2n,n+1)$,
see Fig. \ref{F5}. Provided an avalanche has started, the total
number of particles in the basin of attraction is $n(n+1)+1$.
Again, not all of these particles can be delivered to the target
site $(2n,n+1)$, because, before reaching that site, the avalanche
has to pass through $N_{\rm odd}(2n)$ of intermediate rows with
odd number of sites, leaving a particle at each of them. The new
feature here is that $N_{\rm odd}(2n)$ depends on the parity of
the number $n$, because the intersection of row $n+1$ with the
basin of attraction has the same number of sites as the row $n$.

(i) When $n$ is even $N_{\rm odd}(2n) = n-2$ and \beq h_{\rm
max}(2n,n+1)= n(n+1)+1 -(n-2)=n^2 +3 = \tau^2/4 +3, \quad \tau/2\;
\mathrm{even}. \label{eveneven} \eeq

This case is illustrated for $\tau =8$ by the avalanche shown in
Fig. \ref{F5}

(ii) When $n$ is odd $N_{\rm odd}(2n) = n-1$ and \beq h_{\rm
max}(2n,n+1)= n(n+1)+1 -(n-1)=n^2+2 = \tau^2/4 +2, \quad \tau/2\;
\mathrm{odd}. \label{evenodd} \eeq

\begin{figure}[h]
\center
\includegraphics[width=100mm]{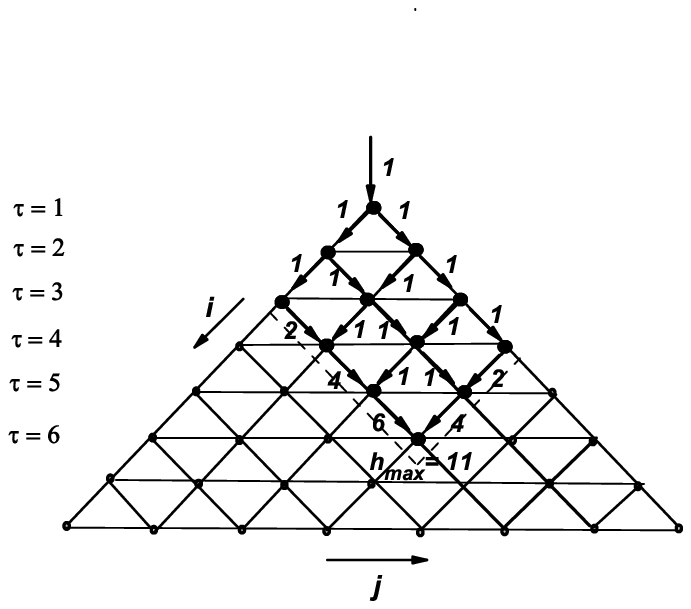}
\caption{The same as in Fig. \ref{F5} for even-$\tau$ row when
$\tau/2$ is odd.}\label{F6}
\end{figure}

This case is illustrated for $\tau =6$ by the avalanche shown in Fig. \ref{F6}.

Consider next a shift by $p=1,2,\dots, n-1$ sites to the left of
the left central site $(2n,n)$ or to the right of the right
central site $(2n,n+1)$. Such a shift changes the basin of
attraction from the rectangle $n\times (n+1)$ to a rectangle
$(n-p)\times (n+p+1)$. Thus, provided an avalanche has started, in
the fully occupied stable configuration of the basins of
attraction of the sites $(2n,n-p)$ and $(2n,n+p+1)$ there are
$(n-p)(n+p+1)+1$ particles. Note that from row $n-p$ up to row
$n+p+1$ there are $2p+2$ segments with equal number of sites
$n-p$. Thus, the number of segments with odd number of sites in
the basin of attraction depends on the parity of $n-p$.

(2a) When $n-p$ is even, all the particles from the $2p+2$ central
even-length segments can be transferred downstream by an
avalanche. Excluding the initial and final sites, there remain
$2(n-p-2)$ intermediate unequal-length segments, half of which
contain odd number of sites. Therefore, the maximum number of
particles that can occupy sites $(2n,n-p)$ and $(2n,n+p+1)$ with
$n-p$ even is $$(n-p)(n+p+1)+1 -(n-p-2)=n^2-p^2+3.$$

(2b) When $n-p$ is odd, all avalanches leave one particle on the
$2p+2$ central odd-length segments. Excluding the initial and
final sites, there remain again $2(n-p-2)$ intermediate
unequal-length segments: $n-p-2$ with length increasing from 2 up
to $n-p-1$ and $n-p-2$ with length decreasing from $n-p-1$ down to
2. Since each of the above series of segments begins and ends with
an even-length segment, the number of intermediate segments with
odd number of sites is $n-p-3$. Therefore, the maximum number of
particles that can occupy sites $(2n,n-p)$ and $(2n,n+p+1)$ with
$n-p$ odd is $$(n-p)(n+p+1)+1 -(n-p-3)-(2p+2)=n^2-p^2-2(p-1).$$

This completes the proof of Proposition II.

\section{Discussion}

Here we have made an attempt to use the directed Abelian algebras,
recently introduced by Alcaraz and Rittenberg \cite{AR}, in the
study of directed avalanches with stochastic toppling rules on the
rotated square lattice. We have considered the directed quadratic
algebra (\ref{oural}) which corresponds exactly to the stochastic
toppling rules of the avalanches analytically studied in \cite{PB}
and \cite{KMT}. Within different continuous approximations, the
latter works predicted a consistent set of critical exponents
which was questioned by the large-scale computer simulations in
\cite{AR}.

We have derived exact expressions for the probabilities
(\ref{Ck2p}) of all possible toppling events which follow the
transfer of arbitrary number of particles  to a site in the
stationary configuration, see (\ref{neven}), (\ref{nodd}). We
have suggested a description (\ref{evolava}) of the virtual-time
evolution of directed avalanches on two dimensional lattices from
which, in principle, the probability distribution of avalanche
durations can be derived. However, the solution of the problem for
large times seems untractable.

We succeeded in applying the algebraic approach only to the
extreme cases of directed deterministic avalanches (when $\alpha =
\beta =0$ in (\ref{oural})) and trivial stochastic avalanches
describing simple random walks of two particles (when $\alpha,
\beta >0$ and $\gamma =0$ in (\ref{oural})). However, the study of
these cases has clarified the role of each particular kind of
toppling in the process of avalanche growth. For example, the
process which ensures both the avalanche growth and decay is the
toppling of an unstable site which transfers odd number of
particles to each of its nearest neighbors ahead: if both of these
neighbors are occupied (empty), the number of particles in the
avalanche increases (decreases) by two. On the other hand, if
particles are transferred to the two neighbors ahead in even
portions, branching of the avalanche occurs without gain or loss
of particles.

In the general case we have determined exactly a number of
important maximum possible values of: the current at any given
odd, (\ref{Imaxodd}), and even, (\ref{Imaxeven}), moment of time; the
occupation number (`height') of each site at any moment of time,
see Proposition II. Our results for the maximum current reveal a
quadratic increase with time $\tau$, with leading asymptotic
behavior $I_{max}(\tau)\propto \tau^2/2$. The leading asymptotic form of the
maximum height is quadratic in time $I_{max}(\tau)\propto \tau^2/4$
at a finite distance from the cental site(s),
while at a finite distance $d \geq 1$ from a
closed boundary it changes to the linear one $h_{max}(\tau)\propto
d\tau$, for $d$ even, and $h_{max}(\tau)\propto (d-1)\tau$ for $d$
odd. The above asymptotic laws have easy heuristic explanation in
terms of number of particles involved in the relevant domain of
sites.

Note that the maxima for the local heights are unconstrained,
hence, in general, they do not happen simultaneously in any
particular avalanche. The maximum height (occupation number) at a
site, at a given virtual moment of time $\tau$, is attained in
particular avalanches which deplete to the maximum possible extent
the basin of attraction of that site and focus the flux of
particles onto it. As a result, in these special avalanches all
the remaining sites do not receive any particles at the moment
$\tau$, hence, they are left stable.

The above extreme values have been established on fully occupied
stable configurations in a region of the lattice, the probability
of which in the stationary state vanishes with the time $\tau$ as
$2^{-\tau^2/2}$ for the maximum current, or $2^{-\tau^2/4}$ for
the maximum height. The statistical weight of such events depends
crucially on the still unsolved problem for obtaining the
probability of all different avalanches at which the given values are
realized.

As far as the temporal dependence of the averaged in the
stationary state of the system avalanche front width $w_{\mathrm
av}(\tau)$ and mean site occupation number $h_{\mathrm av}(\tau)$
are concerned, the existing theories \cite{PB}, \cite{KMT}, as
well as the simple random walk picture \cite{Bu} and computer
simulations \cite{V}, \cite{SV}, agree upon the scaling laws
$w_{\mathrm av}(\tau)\propto \tau^{1/2}$ and $h_{\mathrm av}(\tau)
\propto \tau^{1/4}$. Obviously, these predictions pertain only to
the stage of growth of the avalanches.

\section*{Acknowledgement}

The support of a grant for JINR - Bulgaria collaboration project
is gratefully acknowledged.

\end{document}